\documentclass[a4,12pt, letters,  usenatbib]{mn2e}
\usepackage[dvipdf]{graphicx, psfrag}
\usepackage{times}
\usepackage{color}
\usepackage{amsmath, amssymb}

\usepackage{natbib}
\usepackage{aas_macros}

 \title[Prolate stars due to meridional flows]
 {Prolate stars due to meridional flows}
\author[K. Fujisawa, \& Y.Eriguchi]
{Kotaro Fujisawa
\thanks{E-mail: fujisawa@ea.c.u-tokyo.ac.jp} and
Yoshiharu Eriguchi \\
Department of Earth Science and Astronomy,
Graduate School of Arts and Sciences, University of Tokyo,
Komaba, Meguro-ku, Tokyo 153-8902, Japan
}
\date{Accepted 2013 November 1. Received 2013 November 1; in original from 2013 October 7}
\pubyear{2014}

\def\Vec#1{\mbox{\boldmath $#1$}}

\def\P#1#2{\dfrac{\partial #1}{\partial #2}}
\def\PP#1#2{\dfrac{\partial^2 #1}{\partial #2^2}}

\begin{document}

\maketitle

\begin{abstract}

We have shown analytically that shapes 
of incompressible stars could be prolate
 if appropriate meridional flows exist.
Although this result is strictly valid only if either 
the meridional flow or the rotation is absent
and the vorticity is associated uniformly  
with meridional flow, this implies that perpendicular 
forces against centrifugal 
and/or magnetic forces might play 
important roles within stars.  
A consequence of the presence of meridional 
flows might be to decrease stellar oblateness
due to centrifugal and/or magnetic fields.

\end{abstract}


\section{Introduction}

Stellar shapes have long been considered
to be oblate (including a spherical shape)
due to the effects of centrifugal and/or 
magnetic forces. However, recently 
\cite{Kuhn_et_al_2012} have revealed 
that the shape of
our Sun is `perfectly round' against
the common expectation of an oblate
shape due to its rotation. At present,
no clues have been proposed to solve this
`strange' problem (see e.g. 
\citealt{Gough_2012}).

Concerning stellar deformation,
the effect of magnetic fields has been 
widely investigated (see e.g. 
\citealt{Chandrasekhar_Fermi_1953},
\citealt{Ferraro_1954},
\citealt{Tomimura_Eriguchi_2005}, 
\citealt{Yoshida_Eriguchi_2006}, 
\citealt{Haskell_et_al_2008},
\citealt{Yoshida_Yoshida_Eriguchi_2006},
\citealt{Lander_Jones_2009}, and
\citealt{Fujisawa_Yoshida_Eriguchi_2012}).

The results found thus far are that 
purely poloidal magnetic fields make 
stars oblate, while  purely toroidal 
magnetic fields lead stars to become
prolate. Very recently, 
\cite{Ciolfi_Rezzolla_2013} succeeded 
in obtaining equilibrium states of
magnetized stars with mixed 
poloidal-toroidal magnetic fields,
even for configurations 
with very large toroidal magnetic fields.
They showed that configurations 
with {\it strong  toroidal} magnetic 
fields could be prolate.
Thus, poloidal and toroidal magnetic 
fields act as increasing and decreasing
mechanisms for the stellar oblateness, 
respectively.

However, we should point out that flows 
within stars might work as one deforming 
mechanism of stellar configurations.
In particular, the effect of meridional
flows might make stellar shapes prolate,
as shown in \cite{Eriguchi_Muller_Hachisu_1986}
and \cite{Birkl_Stergioulas_Muller_2011}, 
although they did not describe their
results quantitatively from the point 
of deformation due to meridional flows.

In this Letter, we deal analytically with meridional 
circulations of incompressible stars
with {\it slow flow velocities} 
in order to show  the deformation due to meridional flows 
quantitatively and clearly.

\section{Problem and Solution}

\subsection{Stationary states of axisymmetric barotropic stars}

{\it Axisymmetric barotropic stars} in {\it stationary states} need to satisfy the following condition:
\begin{eqnarray}
 \int {dp \over \rho} = -\phi -{1 \over 2} (v_r^2 + v_{\theta}^2) 
- \int {\omega_{\varphi} \over \rho r \sin \theta} d \psi + \int R \Omega^2 d R  + C \  \ .
\end{eqnarray}
Here $\phi$, $\omega_{\varphi}$, $\psi$ and $C$ are the gravitational potential, the
$\varphi$-component of vorticity, the stream function and an integral constant, 
respectively, and other symbols have their usual meanings. The spherical coordinates $(r, \theta, \varphi)$
are mainly used but sometimes the cylindrical coordinates $(R, \varphi, z)$ are also used for convenience. 
We assume the density is a constant in order to find analytical solutions, i.e
\begin{eqnarray}
 \rho(r,\theta) = \rho_0 ({\rm constant}) \ .
 \end{eqnarray}
We also assume the star has no wind external to itself, i.e
\begin{eqnarray}
 \rho(r, \theta) = 0 ({\rm outside}) \ .
\end{eqnarray}
The stream function is defined by
\begin{eqnarray}
 v_r = \frac{1}{r^2 \sin \theta \rho_0} \P{\psi}{\theta} \ ,
\end{eqnarray}
\begin{eqnarray}
 v_\theta = -\frac{1}{r \sin \theta \rho_0} \P{\psi}{r} \ .
\end{eqnarray}
It should be noted that in this expression the following situations must be taken into account in order to satisfy the integrability condition of the equations of motion:
\begin{eqnarray}
  {\omega_{\varphi} \over \rho_0 R}
  & = & - \nu(\psi) \ , \\
  \Omega & = & \Omega(R) \ ,
\end{eqnarray}
where $\nu(\psi)$ is an arbitrary function of the stream function and the angular velocity $\Omega$ is an arbitrary function of $R$. 

 We need to note that our calculations should
 be used either for purely rotating stars
 or for stars with purely circulating flows
 within meridional planes. The reason for this
 is that we have not taken into account the 
 $\varphi$-component of the equation of 
 motion, i.e.
 \begin{eqnarray}
    \Vec{v} \cdot \nabla (R \Omega) =  0 \ .
 \end{eqnarray}
 As is explained in 
 \cite{Eriguchi_Muller_Hachisu_1986},
 for non-singular angular velocity 
 distributions, this condition results in
 two situations: (i) purely rotating stars,
 i.e. $v_r = 0$ and $v_{\theta} = 0$, or
 (ii) configurations with constant angular
 momentum throughout the whole star, i.e.
 $ \Omega$ needs to vanish to avoid 
 singular behavior of the angular velocity on
 the rotation axis.
 Thus our solutions in this Letter need
 to be used for either rotating stars without
 meridional flows or for non-rotating stars
 with meridional flows.

The stream function must satisfy the following equation:
 \begin{eqnarray}
  \PP{\psi}{r}  +  \frac{\sin \theta}{r^2} \P{}{\theta} 
 \left(\frac{1}{\sin \theta} \P{\psi}{\theta} \right) 
  = - r\sin \theta \rho_0 \omega_\varphi \ .
 \label{Eq:Delta_psi}
 \end{eqnarray}

The distributions of the density, pressure and stream function can be 
obtained from the above stationary condition and the equation for the stream 
function, once the forms of arbitrary functions $\nu(\psi)$ and $\Omega(R)$ and 
the barotropic relation between the density and the pressure
are specified. 

The boundary conditions for the gravitational potential and 
the stream function are as follows: (i) the gravitational potential 
behaves as $1 /r$ at infinity and the steam function is constant 
along the (unknown) stellar surface. Considering these boundary conditions, 
the gravitational potential and the stream function can be expressed by the integral forms as
%
%
\begin{eqnarray}
 \phi(r,\theta) 
 = - 4 \pi G \sum_{n=0}^{\infty} P_{2n}(\cos\theta) 
   \int_0^{\pi/2} d\theta'\sin \theta P_{2n}(\cos\theta') \cr
   \times \int_0^{r_s(\theta')} d r' r'^2  f_{2n}(r,r')
 \rho_0 \ ,
\end{eqnarray}
%
\begin{eqnarray}
 \psi(r,\theta)
 =   r \sin \theta \sum_{n=1}^{\infty}
 \frac{P_{2n-1}^1(\cos\theta)}{2n(2n-1)} 
   \int_0^{\pi/2} d\theta' \sin\theta' \cr 
   \times   P_{2n-1}^1(\cos\theta') 
       \int_0^{r_s(\theta')} d r' r'^2 f_{2n-1}(r, r')
\rho_0 \omega_\varphi(r',\theta') \cr
 + r\sin \theta \sum_{n = 1}^{\infty} \alpha_{2n-1}P_{2n-1}^1(\cos \theta) r^{2n-1}  \ ,
\end{eqnarray}
where 
$P_n$ is the Legendre function and 
 $r_s(\theta)$ expresses the shape of the
deformed surface of the star,
%
%
\begin{eqnarray}
   f_{n}(r,r')
 = \left\{
   \begin{array}{lr}
     r'^{n}/r^{n+1} , & (r \geq r'), \\
     r^{n}/r'^{n+1} , & (r \leq r').
   \end{array}
 \right.  
\end{eqnarray}
%
%
and $\alpha_n$ are coefficients. 
Since we have assumed that there is no external wind,
the $r$ component of the velocity ($v_r$)
must vanish at the surface. 
Therefore, the boundary condition for the 
stream function on the surface is as follows:
\begin{eqnarray}
 \psi(r_s, \theta) = 0.
\end{eqnarray}
We fix the coefficients to 
fulfill this boundary condition 
(\citealt{Eriguchi_Muller_Hachisu_1986}; \citealt{Fujisawa_Eriguchi_2013}).

\subsection{Stationary configurations of
incompressible fluids with very slow flow velocities}

In order to find analytical solutions to the basic equations described above, we further assume the following situation.
(i) The form of the arbitrary function $\nu(\psi)$ is specified as follows:
\begin{eqnarray}
 \nu(\psi ) = \varepsilon \nu_0 ({\rm constant}) \ .
\end{eqnarray}
(ii) The rotational velocity is a constant, 
i.e.,
\begin{eqnarray}
   \Omega = \varepsilon \Omega_0 ({\rm constant}) \ ,
\end{eqnarray}
where $\varepsilon$ is a small constant that expresses 
the slow fluid velocities in both the meridional plane and the $\varphi$-direction.

For an incompressible body, we need only to solve for the surface shape by 
setting $p = 0$ on the surface, instead of solving for 
the density distribution. Thus, the solutions for our problem can be studied by expanding the quantities
with respect to the small quantity $\varepsilon$, as 
\begin{eqnarray}
  r_s(\theta) & = & \sum_{n = 0}^{\infty}\varepsilon^{2n} r_s^{(2n)}(\theta) \ , \\
  \psi(r,\theta) & = & \sum_{n = 0}^{\infty} \varepsilon^{n} \psi^{(n)}(r,\theta) \ ,
\end{eqnarray}
where $r_s^{(2n)}$ and $\psi^{(n)}$ are
corresponding quantities of the surface shape and the stream function, respectively.
Other physical quantities are also
expanded, as
\begin{eqnarray}
  F(r,\theta)  =  \sum_{n = 0}^{\infty} \varepsilon^{n} F^{(n)}(r,\theta) \ ,
\end{eqnarray}
where $F(r,\theta)$ expresses a certain physical
quantity.

For simplicity, we choose the spherical
configurations without flows as the $n = 0$ terms, i.e.
\begin{eqnarray}
   r_s^{(0)}(\theta) =  r_0 ({\rm constant}) \ , 
\end{eqnarray}
\begin{eqnarray}
  \psi^{(0)}(r,\theta) =  0 \ , 
\end{eqnarray}
\begin{eqnarray}
 \frac{dp^{(0)}(r,\theta)}{dr} =  -\rho_0 \frac{d\phi^{(0)}}{dr} \ .
\end{eqnarray} 

The stationary equations are written 
to the second lowest order with respect to
the small quantity $\varepsilon$ as follows:
\begin{eqnarray}
\begin{split}
 \frac{p^{(0)}}{\rho_0}  +  \frac{\varepsilon^2 p^{(2)}}{\rho_0}
=&  - \phi^{(0)}(r) + C^{(0)} \\
+&   \varepsilon^2 \left[ - \phi^{(2)} (r,\theta) - {1 \over 2} ( v_{r}^{(1)2}(r,\theta) + v_{\theta}^{(1)2}(r,\theta) ) \right.  \
  \\ +& \left.  \nu_0 \psi^{(1)}(r,\theta) + {1 \over 2} r^2 \sin^2 \theta \Omega_0^2 + C^{(2)} \right]
 \ , 
\end{split}
\end{eqnarray}
where
\begin{eqnarray}
\begin{split}
 \varepsilon \psi^{(1)}(r,\theta)
& =  - {1 \over 30} \varepsilon 
\rho_0^2  \nu_0 r^2 (5 r_0^2 - 3 r^2)
\sin^2 \theta \\
& + r_0 \sin \theta  \varepsilon
\sum_{n = 1}^{\infty} \alpha_{2n-1}^{(1)}
P_{2n-1}^1(\cos \theta) r^{2n-1}  \ .
\end{split}
\end{eqnarray}

When we apply the boundary condition 
to the above stationary equation
on a {\it deformed surface},
i.e. 
\begin{eqnarray}
p^{(0)}(r_0) + 
\varepsilon^2 {d p^{(0)}(r) \over dr}|_{r = r_0}
r_s^{(2)}(\theta)
+ \varepsilon^2 p^{(2)}(r_0,\theta) = 0  \ ,
\end{eqnarray}
we obtain
\begin{eqnarray}
0 
& = & - \phi^{(0)}(r_0)
 - \varepsilon^2 {d \phi^{(0)}(r) 
\over dr}|_{r = r_0} r_s^{(2)}(\theta) 
+ C^{(0)} \cr
 & + & \varepsilon^2 \left[ - \phi^{(2)} (r_0,\theta)
 - {1 \over 2}( v_{r}^{(1)2}(r_0,\theta) 
+ v_{\theta}^{(1)2}(r_0,\theta) ) \right.  \cr
 & +& \left. \nu_0 \psi^{(1)}(r_0,\theta) 
+ {1 \over 2}r_0^2 \sin^2 \theta \Omega_0^2 + C^{(2)} \right]
 \ .
\end{eqnarray}
Here
\begin{eqnarray}
  \phi^{(0)}(r) 
 =   {2 \pi G \over 3}\rho_0 r^2
- 2 \pi G \rho_0 r_0^2 \ , 
\end{eqnarray}
\begin{eqnarray}
 \psi^{(0)}(r,\theta) =
 r \sin \theta \sum_{n = 1}^{\infty}\alpha_{2n-1}^{(0)} 
P_{2n-1}^1(\cos \theta) r^{2n-1} = 0 \ ,
\end{eqnarray}
\begin{eqnarray}
 \psi^{(1)}(r,\theta)
& = &  - {1 \over 30}  
\rho_0^2  \nu_0 r^2 (-3 r^2 +5r_0^2)
\sin^2 \theta \cr
&  & + r \sin \theta  \sum_{n = 1}^{\infty}
\alpha_{2n-1}^{(1)}P_{2n-1}^1(\cos \theta) r^{2n-1}  
\ .
\end{eqnarray}
Since this quantity $\psi^{(1)}$ is 
the first order term with respect to 
$\varepsilon$, we only need to consider
the boundary condition for the stream function on the {\it undeformed surface}. 
Thus,
\begin{eqnarray}
 \psi^{(1)}(r,\theta) =  {1 \over 10} 
\rho_0^2  \nu_0 \sin^2 \theta 
r^2 (r^2 - r_0^2) 
 \ ,
 \label{Eq:psi} 
\end{eqnarray}
\begin{eqnarray}
  v_r(r,\theta) = {1 \over 5} \rho_0 \nu_0 (r^2-r_0^2) 
\cos \theta \ ,
\end{eqnarray}
\begin{eqnarray}
  v_\theta(r,\theta) = - {1 \over 5} \rho_0 \nu_0 
(2 r^2-r_0^2) \sin \theta \ .
\end{eqnarray}
can be obtained. These can be expressed 
by using $\alpha_{2n-1}^{(k)}$ as follows:
\begin{eqnarray}
 \alpha_{2n-1}^{(0)} & = & 0 \ , \cr
 \alpha_{1}^{(1)} & = & 
  {1 \over 15} \rho_0^2 \nu_0 r_0^2 \ , \cr
 \alpha_{2n+1}^{(1)} & = & 0 \ , 
\end{eqnarray}
where $n = 1, 2, \dots$.

From the second order terms with respect to 
$\varepsilon$, the following condition
can be derived:
\begin{eqnarray}
0 
& = & -{4 \pi G  \over 3} \rho_0 r_0 
  r_s^{(2)}(\theta) \cr   
&   & + 4 \pi G \rho_0 r_0  
\sum_{n=0}^{\infty} P_{2n}(\cos \theta) \cr
&  & \times \int_0^{\pi/2} d\theta'\sin \theta' 
P_{2n}(\cos\theta') r_s^{(2)}(\theta') \cr
&  & - {1 \over 2} v_{\theta}^{(1)2}(r_0,\theta) 
 + {1 \over 2} r_0^2 \sin^2 \theta \Omega_0^2 + C^{(2)} 
 \ .
\end{eqnarray}

By expanding the quantity $r_s^{(2)}(\theta)$
as follows:
\begin{eqnarray}
   r_s^{(2)}(\theta) = \sum_{n = 0}^{\infty}
 \beta_{2n}^{(2)} P_{2n}(\cos \theta) \ ,
\end{eqnarray}
where $\beta_{2n}^{(2)}$ are certain constants,
we obtain the following results for
unknown quantities:
\begin{eqnarray}
 C^{(0)} = -{ 4 \pi G \rho_0 \over 3} r_0^2 \ ,
\end{eqnarray}
\begin{eqnarray}
 C^{(2)} = -{ 8 \pi G \rho_0 \over 3} r_0 
\beta_0^{(2)}
  +{ 1 \over 75}  \rho_0^2 \nu_0^2 r_0^4 - {1 \over 3}r_0^2 \Omega_0^2 \ ,
\end{eqnarray}
\begin{eqnarray}
 \beta_2^{(2)} = {1 \over  8 \pi G}
  \left[ {1 \over 5} \rho_0 \nu_0^2 r_0^3
         - {5 \over \rho_0} r_0 \Omega_0^2 \right]
 \ ,
\end{eqnarray}
\begin{eqnarray}
 \beta_{2n}^{(2)} =  0 \ ( n = 2, 3, \dots) \ . 
\end{eqnarray}
Here if we require some condition for the scale
of the star, we can calculate the value of
$\beta_0^{(2)}$ and complete solutions
for our problem to the second order in
$\varepsilon$ can be obtained.

\subsection{Rotation vs circulation}

From the analysis in the previous 
subsection, the change of the surface
due to circulation and/or the rotation
can be expressed to second order
of a certain small quantity $\varepsilon$
as follows:
\begin{eqnarray}
  r_s(\theta) 
 =  r_0 + \varepsilon^2
\Bigg[ \beta_0
   + {1 \over 8 \pi G}
\left( {1 \over 5} \rho_0 \nu_0^2 r_0^3
      - {5 \over \rho_0} r_0 \Omega_0^2 
\right) P_2(\cos \theta) \Bigg] \ .
\end{eqnarray}
From this equation
\begin{eqnarray}
  {r_{\rm equator} - r_{\rm pole} \over r_0}
 = -{3 \over 16 \pi G}
\left( {1 \over 5} \rho_0 \nu_0^2 r_0^2
      - {5 \over \rho_0} \Omega_0^2 
\right) \ ,
\end{eqnarray}
where $r_{\rm pole}$ and 
$r_{\rm equator}$ are the polar and equatorial
radii, respectively. As is easily seen,
uniformly {\it rotating} configurations
without circulation ($\Omega_0 \ne 0, \nu_0 = 0$) become {\it oblate},
while non-rotating configurations 
with {\it circulation} ($\Omega_0 = 0, \nu_0 \ne 0$) 
are {\it prolate}.
In other words, the presence of meridional
flows acts as an {\it effective force}
perpendicular to the equatorial plane.

\begin{figure*}
 \includegraphics[scale=0.85]{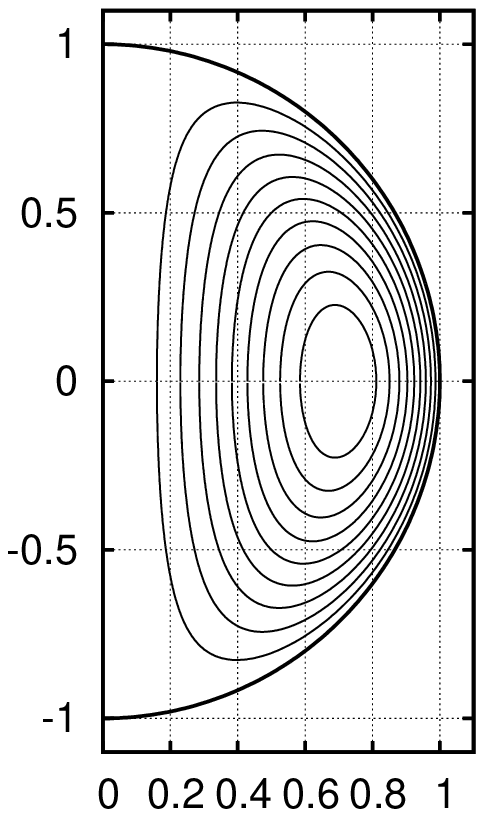}
 \includegraphics[scale=0.85]{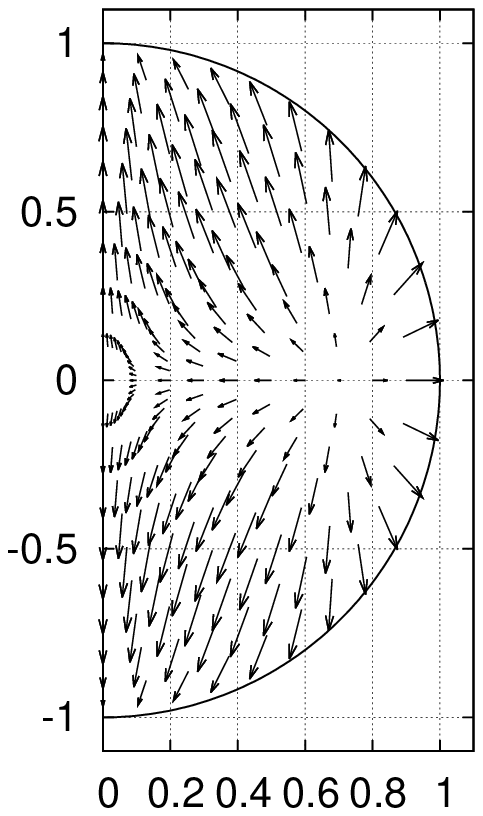}
 \includegraphics[scale=0.85]{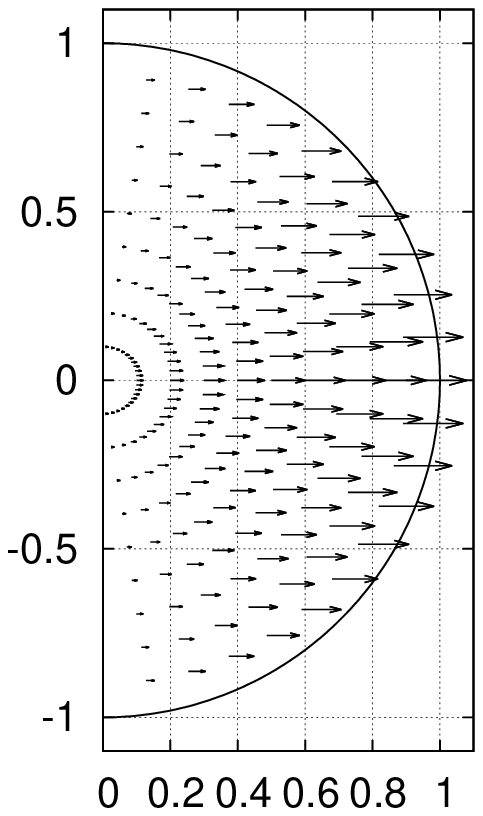}

 \caption{Left: contours of $\psi^{(1)}(r.\theta)$ 
in Eq. (\ref{Eq:psi}). The difference between two 
adjacent contours with lines is 1/10 of the 
maximum value of $\psi^{(1)}(r,\theta)$.
The outermost curve denotes the stellar 
surface.
Centre: the force due to the meridional 
flows (Eq. \ref{Eq:F_mer_r} and Eq. \ref{Eq:F_mer_th}).
Right: the centrifugal force 
of uniform rotation.
 }
 \label{Fig}
\end{figure*}

This can be clearly seen by defining
the effective force due to meridional 
flows with $\Omega_0 = 0$ as
\begin{eqnarray}
\Vec{F} \equiv -\frac{\rho_0}{2}  \nabla |v_r^2 + v_\theta^2| + \rho_0 (\Vec{v} \times \Vec{\omega}) \ ,
\end{eqnarray}
The $r$ and $\theta$ components of 
this force for our incompressible
configurations with $\Omega_0 = 0$ are
\begin{eqnarray}
 F_r = \frac{\nu_0^2 \rho_0^3}{50} \left\{  (-4r^3 + 4r_0^2 r) + (8 r^3 - 6r_0^2 r) \sin^2 \theta  \right\} \ ,
  \label{Eq:F_mer_r}
\end{eqnarray}
\begin{eqnarray}
 F_\theta = \frac{\nu_0 \rho_0^3}{50} \left\{  (4r^3 - 6r_0^2 r) \sin \theta \cos \theta \right\} \ .
  \label{Eq:F_mer_th}
\end{eqnarray}
They are shown in the central panel of 
Fig.\ref{Fig}. In this figure, the
centrifugal force due to uniformly rotating
configurations, i.e. $R \Omega_0^2$, is also shown in the right panel.

In order to estimate 
deformation by meridional flows quantitatively, 
we define the deformation ratio $\Lambda$ as
\begin{eqnarray}
 \Lambda \equiv \left| 
   \frac{r_{\rm c. equator} - r_{\rm c. pole}}
  {r_{\rm r. equator} - r_{\rm r. pole}}
\right|,
\end{eqnarray}
where subscripts `$c$' and `$r$' denote
configurations with meridional circulation
and with rotation, respectively.
For incompressible fluids with {\it very slow
flow velocities},
\begin{eqnarray} 
 \Lambda &=& \frac{1}{25} \frac{r_s^2 \nu_0^2 \rho_0^2}{\Omega_0^2} \nonumber \\
         &=& \frac{v_p^2(r_s, \pi/2)}{r_s^2 \Omega_0^2},
\end{eqnarray}
where $v_p$ is the poloidal velocity, 
defined as $v_p^2 = v_r^2 + v_\theta^2$.
Therefore, the stellar deformation depends on the ratio of
the meridional velocity to the rotational velocity at the 
equatorial surface. When we use the solar radius and rotational velocity,
for example, 
\begin{eqnarray}
 \Lambda &\sim& 1.0 \left(\frac{v_p}{1.81 \times 10^5 \mathrm{cm/s}} \right)^2 \nonumber \\ 
&\times& \left( \frac{r_0}{6.96 \times 10^{10}\mathrm{cm}}\right)^{-2} 
\left(\frac{\Omega_0}{2\pi / 28.0 \mathrm{day}}\right)^{-2}.
  \label{Eq:solar_circulation}
\end{eqnarray}
%
 The {\it commonly believed} 
 solar convective flow or meridional flow velocities
 (e.g. \citealt{Miesch_Living_Reviews};
 \citealt{Nordlund_Stein_Asplund_Living_Reviews}) 
 are a few tens or a hundred times smaller than   
 the meridional flow velocity chosen here,
but those values are not too small to be neglected.
Thus, this simple model shows that the meridional flows have 
small but non-zero influence on stellar deformation 
comparing with stellar rotation, which 
is considered as the most powerful 
deformation mechanism within stars.

\section{Discussion and conclusion}

In this Letter, we have obtained the expression for stellar 
deformation due to meridional flows analytically. 
In order to treat the problem 
analytically, we have assumed stationary 
incompressible stars and imposed a $\psi = 0$ 
boundary condition on stellar surfaces.

We have shown that meridional flows 
make stars prolate
under the conditions of our model.
 This might imply that 
meridional flows within stars work
to decrease the oblateness of rotating stars. 
Although we have assumed incompressible fluids,
the role of meridional flows would not
disappear for compressible stars. 
As explained in the Introduction, 
according to the recent very accurate 
observation by \cite{Kuhn_et_al_2012}, 
the solar oblateness is unexpectedly smaller 
than the theoretical value, which is derived
by considering only rotation 
(\citealt{Armstrong_Kuhn_1999}).
\cite{Gough_2012} argued that magnetic fields and/or 
stresses due to turbulence could be possible
mechanisms causing the small oblateness.
However, they did not consider the influence of
meridional flows, which could be one of the possible
mechanisms, as we have shown in this Letter.

One might argue that the velocities of
solar meridional flows are believed to be
much smaller than the values required to
reduce the rotational effect, as seen
from Eq. (\ref{Eq:solar_circulation}).
This belief has been based on 
%
theoretical analysis of 
many observational data on {\it surface} phenomena
(\citealt{Zhao_et_al_2012}). 
On the other hand, we can rely on the theories that
are used to `estimate' physical quantities within
the solar interior. One approach 
for us is to take even the 
`curious' observational data seriously
and to consider possible mechanisms 
thoroughly, even though those mechanism might seem 
to be far from the {\it widely believed} 
theories and/or values.
The `perfectly round' Sun might be
offering us an important hint about 
its interior.

\section*{ACKNOWLEDGEMENTS}

The authors would like to thank the anonymous reviewer for useful
comments and suggestions that helped us to improve this paper. 
This work was supported by Grant-in-Aid for JSPS Fellows.

\bibliographystyle{mn}


\end{document}